\documentclass[aps,pre,twocolumn,superscriptaddress,amsmath,amssymb,nofootinbib]{revtex4-2}

\usepackage{amsmath}
\usepackage{graphicx}
\usepackage{amssymb}
\usepackage{dcolumn}
\usepackage{bm}
\usepackage{color}
\usepackage{float}
\usepackage{upgreek}
\usepackage{gensymb}
\usepackage[final]{hyperref} % adds hyper links inside the generated pdf file
\hypersetup{
	colorlinks=true,       % false: boxed links; true: colored links
	linkcolor=blue,        % color of internal links
	citecolor=blue,        % color of links to bibliography
	filecolor=magenta,     % color of file links
	urlcolor=blue
}
\usepackage{upgreek}
\usepackage[normalem]{ulem}

\makeatletter

\begin{document}
\title{Inertial effects on rectification and diffusion of active Brownian particles in an asymmetric channel}
	
\author{Narender Khatri}\thanks{Corresponding author: narender.khatri@utoronto.ca}
\affiliation{Chemical Physics Theory Group, Department of Chemistry, University of Toronto, Toronto, Ontario M5S 3H6,
Canada}

\author{Raymond Kapral}\thanks{Corresponding author: r.kapral@utoronto.ca}
\affiliation{Chemical Physics Theory Group, Department of Chemistry, University of Toronto, Toronto, Ontario M5S 3H6,
Canada}

\date{\today}
	
\begin{abstract}

Micro- and nano-swimmers moving in a fluid solvent confined by structures that produce entropic barriers are often described by overdamped active Brownian particle dynamics, where viscous effects are large and inertia plays no role. However, inertial effects should be considered for confined swimmers moving in media where viscous effects are no longer dominant. Here, we study how inertia affects the rectification and diffusion of self-propelled particles in a two-dimensional asymmetric channel. We show that most of the particles accumulate at the channel walls as the masses of the particles increase. Furthermore, the average particle velocity has a maximum as a function of the mass, indicating that particles with an optimal mass $M^{*}_{\rm op}$ can be sorted from a mixture with particles of other masses. In particular, we find that the effective diffusion coefficient exhibits an enhanced diffusion peak as a function of the mass, which is a signature of the accumulation of most of the particles at the channel walls. The dependence of $M^{*}_{\rm op}$ on the rotational diffusion rate, self-propulsion force, aspect ratio of the channel, and active torque is also determined. The results of this study could stimulate the development of strategies for controlling the diffusion of self-propelled particles in entropic ratchet systems.

\end{abstract}
	
\maketitle

\section{Introduction}

Many biological microorganisms, as well as artificial active particles, take free energy from their environments and convert it under nonequilibrium conditions into persistent motion. The mechanisms that underlie such active motion and the dynamical properties of these systems are diverse and have been studied extensively~\cite{LP2009,Ramaswamy@ARCMP:2010, Romanczuk_et_al@EPJST:2012,Cates@RPP:2012,2013perspective,A13,Wang2015,Elgeti_et_al@RPP:2015, Zoettl2016,Bechinger_et_al@RMP:2016, Illien2017,  Archer_et_al@AS:2018, GK19, Gaspard_Kapral@Research:2020,Gompper_et_al@JPCM:2020}.

For the most part, the biological and synthetic active agents mentioned above have micrometer or sub-micrometer dimensions and move in viscous environments under conditions where inertia does not play an important role. In such circumstances, the active dynamics is often described by the overdamped Langevin or continuum  models that neglect inertia. Inertia cannot always be neglected, and an increasing body of research~\cite{Lowen-per2020} considers the effects of inertia on active particle motion and describes the new phenomena that arise as a result of its inclusion. While the systems where inertial effects are important are diverse, some examples include systems that support a temperature gradient across coexisting phases~\cite{mandal2019}, vibrobots~\cite{Scholz2016,Torres2016,Scholz_et_al@NC:2018}, active particle motion in low-density media, such as gases~\cite{Sprenger_et_al@PRE:2021,Caprini_et_al@PCCP:2022,Leoni_et_al@PRR:2020}, plasmas~\cite{Morfill_Ivlev@RMP:2009, Arkar_et_al@Molecules:2021, Li_et_al@arxiv:2022}, superfluids~\cite{Kolmakov_Aranson@PRR:2021}, and active aerosols~\cite{Rohde_et_al@OE:2022}, etc.
Such inertia-dominated active particles are termed micro- and nano-flyers rather than swimmers~\cite{Lowen-per2020}.

The rectification of artificial active particles in confined environments in the absence of external forces has attracted interest \cite{Ghosh_et_al@PRL:2013, Ao_et_al@EPJST:2014, Li_et_al@PRE:2014, Ai_et_al@JCP:2014, Reichhardt_Reichhardt@ARCMP:2017, Bisht_Marathe@PRE:2020}. Geometrical confinement controls the volume of phase space that is accessible to the active particles, resulting in entropic barriers that significantly influence their transport properties~\cite{Zwanzig@JCP:1992, Reguera_Rubi@PRE:2001, Kalinay_Percus@PRE:2006,Khatri_Burada@PRE:2020, Khatri_Burada@PRE:2021,Malgaretti_Stark@JCP:2017,Caprini_et_al@PRR:2020,Caprini_et_al@JCP:2019}. As well, confined environments possessing spatial ratchet asymmetry give rise to an entropic ratchet potential that can induce active directed transport in the system \cite{Ghosh_et_al@PRL:2013, Reichhardt_Reichhardt@ARCMP:2017}.

We investigate the underdamped dynamics of self-propelled particles confined by a two-dimensional asymmetric channel. We use a minimal underdamped Langevin model for the dynamics of the self-propelled particles that accounts for inertia. The collisional dynamics of particles with the channel walls are modeled by sliding-reflecting boundary conditions \cite{Reichhardt_Reichhardt@ARCMP:2017,Ghosh_et_al@PRL:2013,Khatri_Burada@PRE:2022}. We focus on how inertial effects influence the rectification and diffusion of active particles in the asymmetric channel.

The article is organized as follows: in Sec.~\ref{sec: Model}, we introduce the underdamped Langevin model used to describe the dynamics of the active particles in the two-dimensional asymmetric channel. Section~\ref{Sec: Distributions} discusses the effects of inertia on the spatial distribution of particles, while Sec.~\ref{Sec: Transport_Properties} presents results on the rectification and effective diffusion in the channel. The main conclusions of the article are given in Sec.~\ref{Sec: Conclusions}.

\section{Model}
\label{sec: Model}

\begin{figure}[hbt]
\centering
\resizebox{1.0\columnwidth}{!}{%
\includegraphics[scale = 1.5]{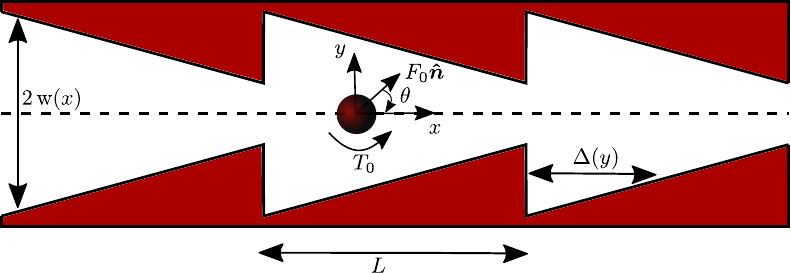}}
\caption{Schematic illustration of an active (self-propelled) Brownian particle of mass $M$ and moment of inertia $I$ confined in a two-dimensional triangular channel with periodicity $L$.
The active force $\boldsymbol{F_0} = F_0 \boldsymbol{\hat{n}}$, angle $\theta$, active torque $T_0$, local width of the channel $2 \, {\rm w} (x) $, and local length of a cell of the channel $\Delta (y)$ are indicated. The shape of the channel structure is prescribed by Eq. (\ref{eq:wall}). The particle cannot penetrate through the channel walls, which are considered rigid; however, the particle is free to rotate and slide along the walls.
}
\label{fig:Model}
\end{figure}

We consider a system comprising active particles dilutely dispersed in a dissipative medium and confined to a two-dimensional asymmetric channel with periodicity $L$ (see Fig.~\ref{fig:Model}). The active particle density is assumed to be sufficiently small so that direct interactions among the particles may be neglected.  As in other studies~\cite{Lowen-per2020}, a minimal underdamped Langevin model is used to describe the active dynamics under conditions where inertia is important. The coupled Langevin equations for an active particle with position $\boldsymbol{r}  = (x, y)$, orientation $\boldsymbol{\hat{n}} = (\cos{ \theta }, \sin{\theta })$, mass $M$, and moment of inertia $I$ read
\begin{equation}\label{Eq:Langevin1}
 \begin{aligned}
    M \ddot{\bm{r}}(t) &= -\gamma_t \Dot{\bm{r}}(t) + F_0 \boldsymbol{\hat{n}} (t) + \gamma_t \sqrt{2 D_t}~ \boldsymbol{\xi} (t), \\
    I \Ddot{\theta}(t) &= -\gamma_r \Dot{\theta}(t) + T_0 + \gamma_r \sqrt{2 D_r} ~ \zeta (t).
    \end{aligned}
\end{equation}
Here $\gamma_t$ and $\gamma_r$ are the translational and rotational friction coefficients, $D_t$ and $D_r$ are the translational and rotational diffusion constants, and $F_0$ and $T_0$ are the active force and torque, respectively. The random variables $\boldsymbol{\xi} (t)$ and $\zeta (t)$ are the Gaussian white noise terms with zero mean and unit variances given by $\langle \boldsymbol{\xi} (t) \otimes \boldsymbol{\xi} (t')\rangle = \delta (t - t') \boldsymbol{\mathrm{1}}$ and $\langle \zeta (t) \zeta (t') \rangle = \delta (t - t')$, respectively, where $\boldsymbol{\mathrm{1}}$ is the unit matrix.
Such a set of coupled Langevin equations can serve as models for a variety of active systems, including systems subject to athermal noise~\cite{Berg@Book:2004,Fily_Marchetti@PRL:2012,Palacci_et_al@Science:2013,Marchetti_et_al@COCIS:2016}, if $\{\gamma_t, \gamma_r, D_t, D_r \}$ are regarded as independent parameters, and $D_t$ and $D_r$ control the strength of noise terms.
For systems with thermal noises that satisfy the fluctuation-dissipation relation, the Einstein relations apply, $D_t = k_B T/\gamma_t$ and $D_r = k_BT/\gamma_r$. The set of coupled Langevin equations~(\ref{Eq:Langevin1}) is a simplified version of a more general set of coupled Langevin equations that applies to asymmetric particles and accounts for translation-rotation coupling~\cite{Martins2022}. The set of general underdamped coupled Langevin equations for chemically-active colloids has been derived using fluctuating chemohydrodynamics~\cite{GK18} and molecular theory~\cite{RSGK20}, where expressions for the active force and torque are given.

The two-dimensional asymmetric and spatially periodic channel shown in Fig.~\ref{fig:Model} is specified as follows: for the upper wall, we have
\begin{equation}\label{eq:wall}
{\rm w}_u (x) = \begin{cases}
{\rm w}_\mathrm{min}, & x = 0\\
{\rm w}_\mathrm{max} - ({\rm w}_\mathrm{max} - {\rm w}_\mathrm{min}) \frac{x}{L}, & 0 < x\leq L,
\end{cases}
\end{equation}
where $L$ is the periodicity of the channel, and ${\rm w}_\mathrm{min}$ and ${\rm w}_\mathrm{max}$ refer to the minimum and maximum half-widths of the channel, respectively.
The dimensionless aspect ratio of the channel is $\epsilon = {\rm w}_\mathrm{min}/{\rm w}_\mathrm{max}$, where we set ${\rm w}_\mathrm{max} = 1$ throughout the work.
The local length of a cell of the channel along $x$ for a given $y$ is $\Delta (y) = ({\rm w}_\mathrm{max} - |y|)L/({\rm w}_\mathrm{max} - {\rm w}_\mathrm{min})$, for ${\rm w}_\mathrm{min} < |y| \le {\rm w}_\mathrm{max}$, and $\Delta (y) =L$, for $-{\rm w}_\mathrm{min} \le y  \le {\rm w}_\mathrm{min}$. For a given value of $x$,  $y$ is bounded between the lower and upper walls. Due to the symmetry about the principal axis of the channel, the lower wall is described by ${\rm w}_l (x) = - {\rm w}_u (x)$. Consequently, $2 \, {\rm w} (x) = {\rm w}_u (x) - {\rm w}_l (x)$ corresponds to the local width of the channel.

When a particle encounters a channel wall, it is elastically reflected~\cite{Reichhardt_Reichhardt@ARCMP:2017, Khatri_Burada@JCP:2019, Khatri_Burada@JSM:2021, Khatri_Burada@PRE:2022}, and its orientation $\theta$ is unchanged during the collision (sliding-reflecting boundary conditions~\cite{Ghosh_et_al@PRL:2013, Khatri_Burada@PRE:2022, Zhang_et_al@PF:2010}).
So, the particle slides along the channel walls until a fluctuation in the orientation vector $\hat{\bm{n}}$ causes it to change so that it may move away from the wall. While we restrict our considerations to this previously-used boundary condition, other boundary conditions that depend on the scale of the wall roughness may be applied. For example, if the particle interacts with the channel walls through rough sphere collisions, the orientation of the particle will change through this collision mechanism, as well as through interactions with the solvent. Consequently, the primary effects on the particle dynamics will likely be more rapid reorientation, decreased residence time near the wall, and reduced rectification.

For translational motion described by Eqs.~(\ref{Eq:Langevin1}), inertial effects dominate frictional effects for times $t \ll M/\gamma_t \equiv \tau_v$, while for orientational motion, they are important for times $t \ll I/\gamma_r \equiv \tau_\omega$, where $\tau_v$ and $\tau_\omega$ are the characteristic times for linear and angular velocity relaxation, respectively. For active particles in a bulk medium, these times may be compared to the reorientation time $\tau_r = 1/D_r$ that determines the time scale on which orientation vector $\hat{\bm{n}}$ decays, and the time $\tau_a = 2a/v_0$ it takes a particle with speed $v_0 = F_0/\gamma_t$ to move a distance equal to the particle diameter $2a$.

For the present study, where the dynamics takes place in a periodic channel, we are interested in the effects of inertia on time scales that reflect the motion on the length scale $L$ of the periodic channel. We define the characteristic diffusion time $\tau=L^2/D_t$, which gauges the time the particle takes to diffuse one period of the channel length. For systems with thermal noise, this characteristic diffusion time is related to $\tau_v$ by $\tau_v/\tau=\big(\tau_v/\tau_{\rm th}\big)^2$, where $\tau_{\rm th}=L/v_{\rm th}$ with $v_{\rm th} =\sqrt{k_B T/M}$ the thermal speed.

Given these considerations, we use a dimensionless description where lengths are scaled by the periodicity of the channel $L$, $\bm{r}' = \boldsymbol{r}/L$, and time by $\tau $, $t'=t/\tau$, analogous to that in Refs.~\cite{SLHS2019,Khatri_Burada@PRE:2021}, so that Eqs.~(\ref{Eq:Langevin1}) read,
\begin{equation}\label{Eq:Langevin2}
 \begin{aligned}
    M^* \Ddot{\bm{r}}(t) &= -{\Dot{\bm{r}}}(t) + f_0 \boldsymbol{\hat{n}} (t) + \sqrt{2} ~\boldsymbol{\xi} (t), \\
    I^* \Ddot{\theta}(t) &= -\Dot{\theta}(t) + t_0 +  \sqrt{2 \alpha} ~ \zeta (t),
    \end{aligned}
\end{equation}
and we dispensed with the primes in writing this coupled set of equations. The dimensionless mass and moment of inertia are given by $M^* = \tau_v/\tau = M D_t /(\gamma_t L^2)$ and $I^* = \tau_\omega/\tau = I D_t/(\gamma_r L^2)$, respectively. In these variables, the dimensionless mass $M^*$ depends on physical mass $M$ as well as $D_t$, $\gamma_t$, and $L$. Similarly, $I^*$ depends on the moment of inertia $I$ as well as $D_t$, $\gamma_r$, and $L$. The dimensionless active force and torque are $f_0 = F_0 L/(D_t \gamma_t)$ and $t_0 =  T_0 L^2/(D_t \gamma_r)$, respectively, and the parameter $\alpha = D_r \tau$. Other choices of dimensionless units could have been used~\cite{Lowen-per2020}.

In the following sections, we present results for the average velocity, effective diffusion coefficient, and other properties obtained from simulations of the coupled Langevin equations~(\ref{Eq:Langevin2}) in the channel ~\cite{note:sim}. At $t = 0$, the particles are uniformly distributed with random orientations in a periodic cell of the channel located between $x = 0$ and $x = 1$. The results are obtained from averages over $10^4$ stochastic trajectories.\\

\section{Spatial distribution}\label{Sec: Distributions}

 \begin{figure}[htb!]
\centering
\resizebox{1.0\columnwidth}{!}{%
\includegraphics[scale = 1.7]{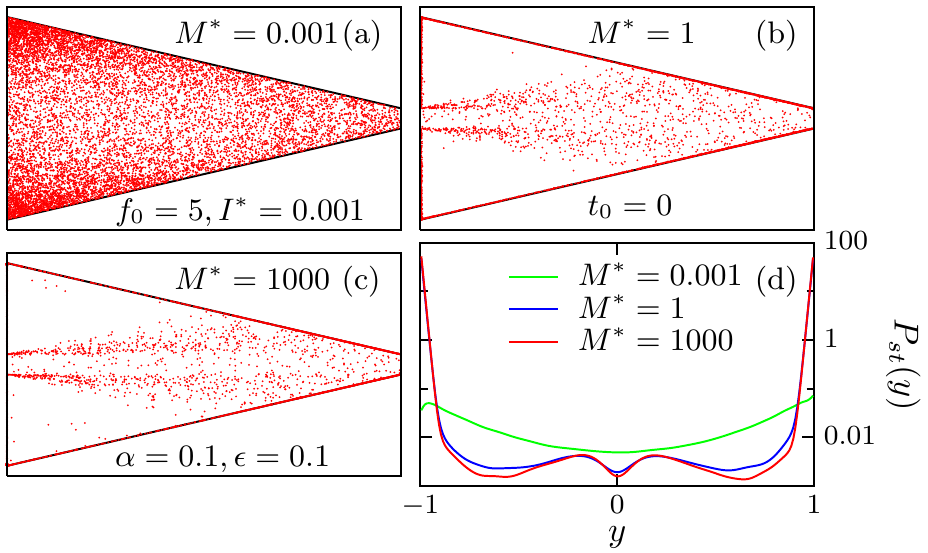}}
\caption{The steady state distribution of particles, mapped into a single cell of the channel, is depicted in (a)-(c) for different values of $M^*$.
The corresponding probability densities $P_{st} (y)$ along the $y$ direction are depicted in (d).
The parameters are: $I^* = 0.001, f_0 = 5, t_0 = 0, \alpha = 0.1$, and $\epsilon = 0.1$.}
\label{fig:Distribution_Graph1}
\end{figure}

In order to analyze the effects of inertia on the rectification and diffusion of self-propelled particles in an asymmetric channel, we first consider the spatial distribution of particles mapped onto a single cell of the channel. Taking $P(x,y)$ to be the probability density of finding a particle at $(x, y)$ in the cell, we define the probability density to find a particle at $y$ per unit local cell length $\Delta(y)$ at this $y$-value: $P(y)=\int_0^{\Delta(y)} dx \;P(x,y)/\Delta(y)$. Figure~\ref{fig:Distribution_Graph1} shows the steady state distribution of particles and the corresponding probability density $P_{st} (y)$ along the $y$ direction for different values of $M^*$. For $M^* \to 0$ in the overdamped limit, the distribution of particles is inhomogeneous, and most of the particles tend to accumulate near the left corners of the cell (see Fig.~\ref{fig:Distribution_Graph1}(a) and the $P_{st} (y)$ plots in (d)). The inhomogeneous distribution of particles can be ascribed to the active motion due to broken detailed balance and the presence of the spatial asymmetry imposed by the shape of the channel~\cite{Ghosh_et_al@PRL:2013, Reichhardt_Reichhardt@ARCMP:2017}.
It is interesting to see that on increasing $M^*$ further, most of the particles quickly accumulate at the channel walls; while the rest of the particles adopt the shape of a funnel about the principal axis of the channel at the middle region due to narrow bottleneck openings (see Fig.~\ref{fig:Distribution_Graph1}(b)).
Most of the particles accumulate at the left corners of the cell, and the distribution of particles is symmetric about the principal axis of the channel (see the $P_{st} (y)$ plots, especially the curve for $M^* = 1$).
Such an observation provides evidence that the rectification of particles in an asymmetric channel can be enhanced by increasing $M^*$.
For strongly underdamped situations where $M^* \to \infty$, the qualitative behavior of the distribution of particles and the corresponding $P_{st} (y)$ are very similar; however, the rectification and diffusion of particles approach zero because inertia dominates the self-propulsion mechanism~\cite{note:largeM}.

\begin{figure}[htb!]
\centering
\resizebox{1.0\columnwidth}{!}{%
\includegraphics[scale = 1.25]{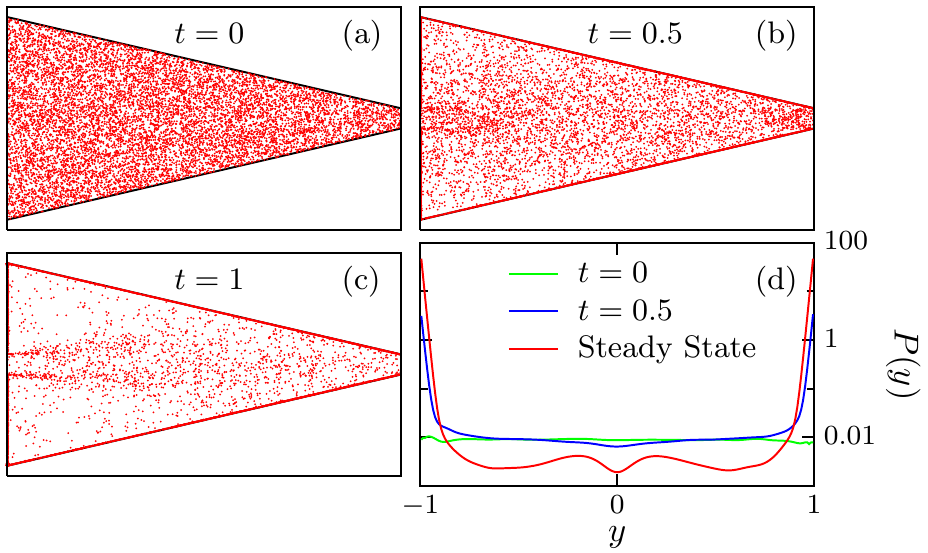}}
\caption{The time evolution of the distribution of particles with $M^* = 1$ is shown in (a)-(c) at increasing times, $t=0, 0.5, 1$, respectively. The probability density $P(y)$ versus $y$ is plotted in (d).
The parameters are: $I^* = 0.001, f_0 = 5, t_0 = 0, \alpha = 0.1$, and $\epsilon = 0.1$.}
\label{fig:Distribution_Graph2}
\end{figure}

The probability density evolution to its steady state distribution is shown in Fig.~\ref{fig:Distribution_Graph2} for a system with $M^* = 1$ and parameters as in Fig.~\ref{fig:Distribution_Graph1}. At $t = 0$, particles are distributed uniformly inside the periodic cell consistent with thermodynamic equilibrium~\cite{Khatri_Burada@JCP:2019, Jacobs@Book:1967, Zwanzig@JCP:1992, Reguera_Rubi@PRE:2001, Kalinay_Percus@PRE:2006}. As time increases to $t=0.5$, the distribution becomes inhomogeneous: particles near the channel walls quickly begin accumulating there, and the funnel-shaped distribution of particles in the middle region of the cell starts to develop. At a yet later time $t=1$, the funnel-shaped distribution sharpens, and particles have continued to accumulate at the walls. The distribution at this time closely resembles the steady state distribution shown in Fig.~\ref{fig:Distribution_Graph1}(b). The probability of accumulation at the left corners of the cell is much higher than in other regions, which is reflected in the structure of the probability density $P(y)$ in Fig.~\ref{fig:Distribution_Graph2}(d).

\section{Rectification and effective diffusion}\label{Sec: Transport_Properties}

\begin{figure}[htb!]
\centering
\resizebox{0.8\columnwidth}{!}{%
\includegraphics[scale = 1.0]{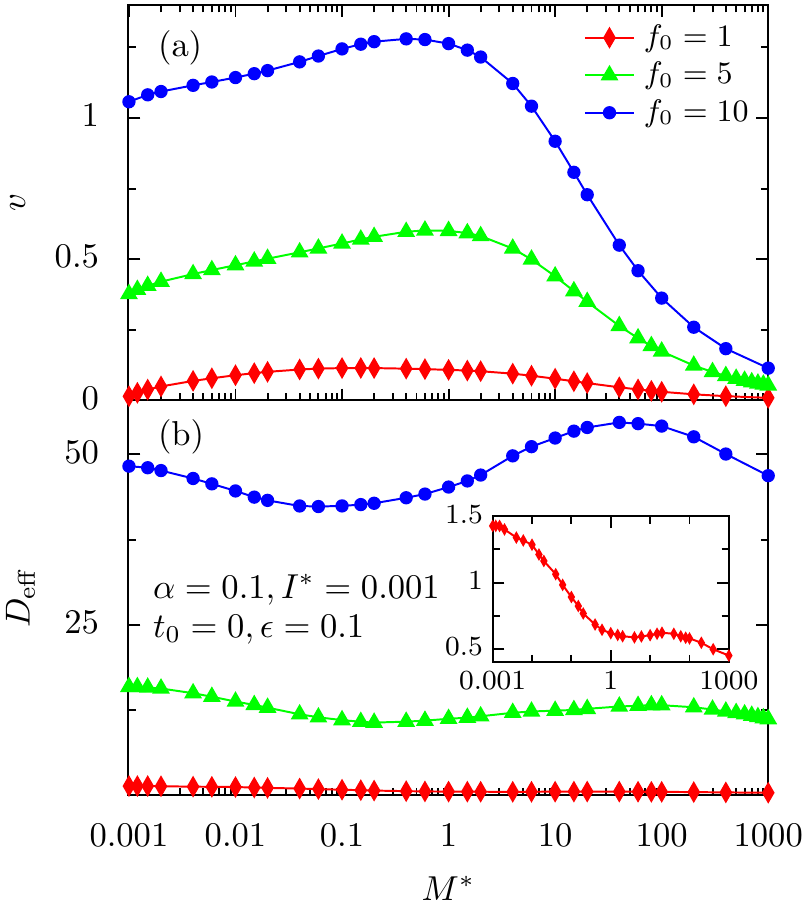}}
\caption{Average velocity $v$ as a function of $M^*$ is shown in (a) for different values of the self-propulsion force $f_0$. The corresponding effective diffusion coefficient $D_{\rm eff}$ is shown in (b).
The inset plots $D_{\rm eff}$ versus $M^*$ for $f_0 = 1$ on an expanded scale to show its structure.
Here and below, the solid lines are guides to the eye, and the statistical errors for $v$ and $D_{\rm eff}$ are smaller than the symbol sizes. The other parameters are: $I^* = 0.001, t_0 = 0, \alpha = 0.1$, and $\epsilon = 0.1$.}
\label{fig:Transport_Properties_Graph1}
\end{figure}

Figure~\ref{fig:Transport_Properties_Graph1} shows the dependence of the average velocity $v$ and effective diffusion coefficient $D_{\rm eff}$ on $M^*$ for different values of the self-propulsion force $f_0$.
We observe that the particles exhibit rectification ($v \neq 0$) in the positive $x$ direction; $v$ is positive due to the chosen shape of the channel. If the shape of the channel were inverted with respect to the $y$ axis, then the magnitude of rectification would remain the same, but $v$ would lie in the negative $x$ direction.
In the small $M^*$ limit, the channel asymmetry affects the particle dynamics more strongly since most of the particles accumulate at the channel walls (see Figs.~\ref{fig:Distribution_Graph1} and \ref{fig:Distribution_Graph2}), resulting in an increase in $v$ with $M^*$. In the strongly underdamped limit, when $M^* \to \infty$, inertia dominates self-propulsion resulting in a rapid decay of $v$ and $D_{\rm eff}$ with $M^*$. Therefore, as one might expect, $v$ has a peak at an optimal mass $M_{\rm op}^*$; thus, in a mixture, rectified particles with $M_{\rm op}^*$ will have a higher speed compared to particles with other $M^*$ values. The peak is more pronounced as $f_0$ increases, and the value of the optimal mass can be controlled by changing $f_0$. For the effective diffusion coefficient, we see that $D_{\rm eff}$ initially decreases with increasing $M^*$, but on increasing $M^*$ further, $D_{\rm eff}$ exhibits an enhanced diffusion peak, which is a signature of the accumulation of most of the particles at the channel walls~\cite{Khatri_Burada@PRE:2022}. Note that the two maxima, in $v$ and in the enhanced diffusion peak, do not occur at the same $M^*$ value.
As expected, $D_{\rm eff}$ increases monotonically with $f_0$. As $f_0 \to 0$, the motion of self-propelled particles approaches that of passive Brownian motion; thus, $v$ will tend to zero, and the enhanced diffusion peak will no longer be present (see the inset of Fig.~\ref{fig:Transport_Properties_Graph1}(b)).

\begin{figure}[htb!]
\centering
\resizebox{0.8\columnwidth}{!}{%
\includegraphics[scale = 1.0]{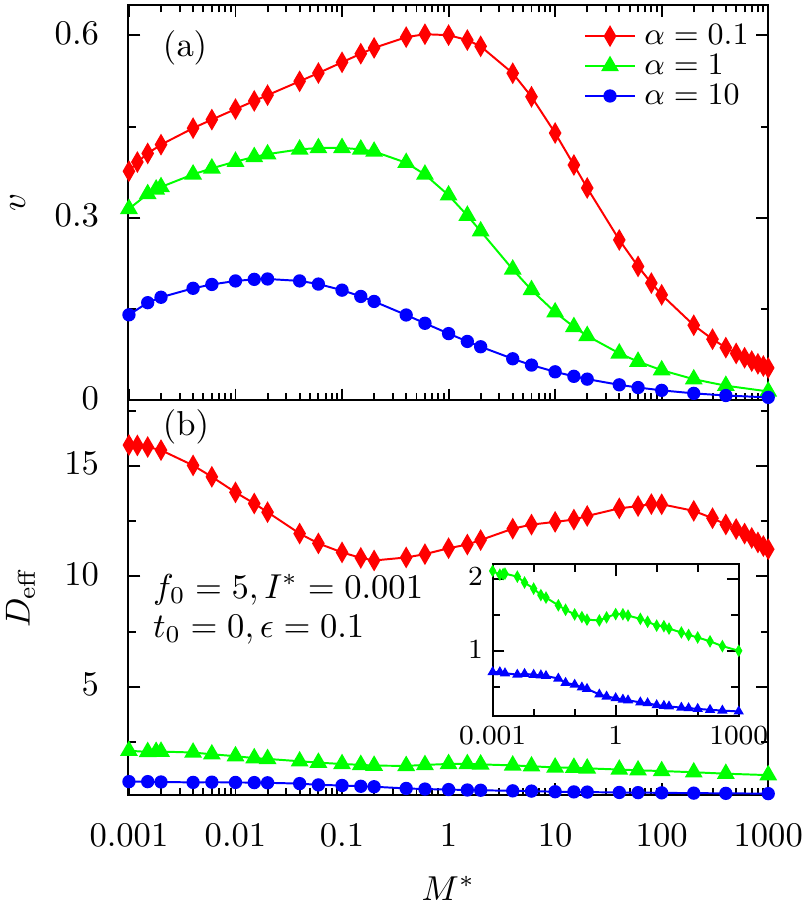}}
\caption{Average velocity $v$ and effective diffusion coefficient $D_{\rm eff}$ as a function of $M^*$ are plotted in (a) and (b), respectively, for different values of the scaled rotational diffusion rate $\alpha$.
The inset in (b) shows $D_{\rm eff}$ versus $M^*$ for $\alpha = 1$ and $\alpha  = 10$ on an expanded scale.
The other parameters are: $I^* = 0.001, t_0 = 0, f_0 = 5$, and $\epsilon = 0.1$.}
\label{fig:Transport_Properties_Graph2}
\end{figure}

\begin{figure}[htb!]
\centering
\resizebox{0.8\columnwidth}{!}{%
\includegraphics[scale = 1.0]{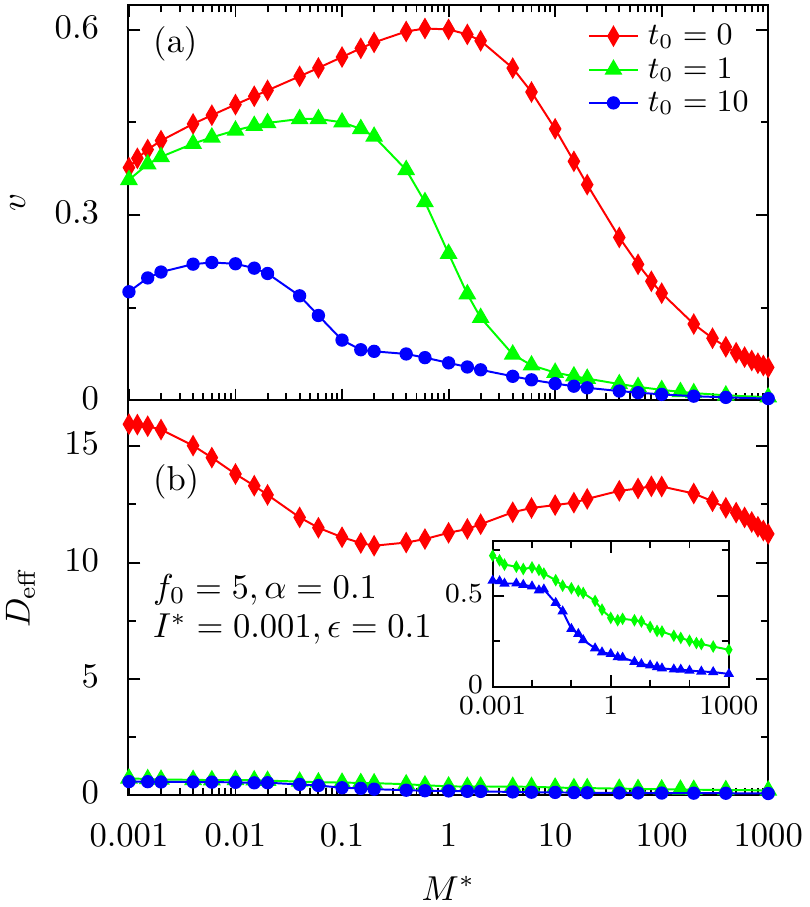}}
\caption{Average velocity $v$ and effective diffusion coefficient $D_{\rm eff}$ as a function of $M^*$ are plotted in (a) and (b), respectively, for different values of the active torque $t_0$.
The inset in (b) shows $D_{\rm eff}$ versus $M^*$ for $t_0 = 1$ and $t_0  = 10$ on an expanded scale.
The parameters are: $I^* = 0.001, \alpha = 0.1, f_0 = 5$, and $\epsilon = 0.1$.}
\label{fig:Transport_Properties_Graph3}
\end{figure}

The variation of $v$ and $D_{\rm eff}$ with $M^*$ for different values of the rotational diffusion rate $\alpha$ is shown in Fig.~\ref{fig:Transport_Properties_Graph2}. The qualitative trends for different $\alpha$ are the same as those described above for different $f_0$; however, the peak is more pronounced as $\alpha$ decreases, and now $M_{\rm op}^*$ can be changed by tuning $\alpha$. As expected, $D_{\rm eff}$ decreases monotonically with increasing $\alpha$. In particular, as $\alpha \to \infty$ where reorientation is rapid, the self-propelled motion tends to passive Brownian motion, $v$ tends to zero, and the enhanced diffusion peak vanishes (see the inset of Fig.~\ref{fig:Transport_Properties_Graph2}(b)).

Figure~\ref{fig:Transport_Properties_Graph3} shows $v$ and $D_{\rm eff}$ versus $M^*$ for different values of the active torque $t_0$. The distribution of particles along the $y$ direction does become asymmetric with respect to the principal axis of the channel when $t_0 \ne 0$, as reported in Refs.~\cite{Khatri_Burada@PRE:2022, Xue@PhD_Thesis:2014,Hargus_et_al@PRL:2021}. For $t_0 > 0$, the density of particles will be higher above the principal axis of the channel; whereas, for $t_0 < 0$, the same density increase will occur in the regime below the principal axis of the channel. However, our channel is symmetric about its principal axis, i.e., it has top-down symmetry, and we have verified that $v$ and $D_{\rm eff}$ are independent of the sign of the active torque $t_0$. Again one observes that the qualitative trends are the same as those discussed above, and the values of $M_{\rm op}^*$ can also be changed by changes in the active torque $t_0$.

Finally, we point out that the qualitative behavior of $v$ and $D_{\rm eff}$ versus $M^*$ remains the same for different values of the moment of inertia $I^*$ and aspect ratio of the channel $\epsilon$. In addition, $M_{\rm op}^*$ is found to be independent of $I^*$, and it is only weakly dependent on $\epsilon$.

\begin{figure}[htb!]
\centering
\resizebox{1.0\columnwidth}{!}{%
\includegraphics[scale = 1.3]{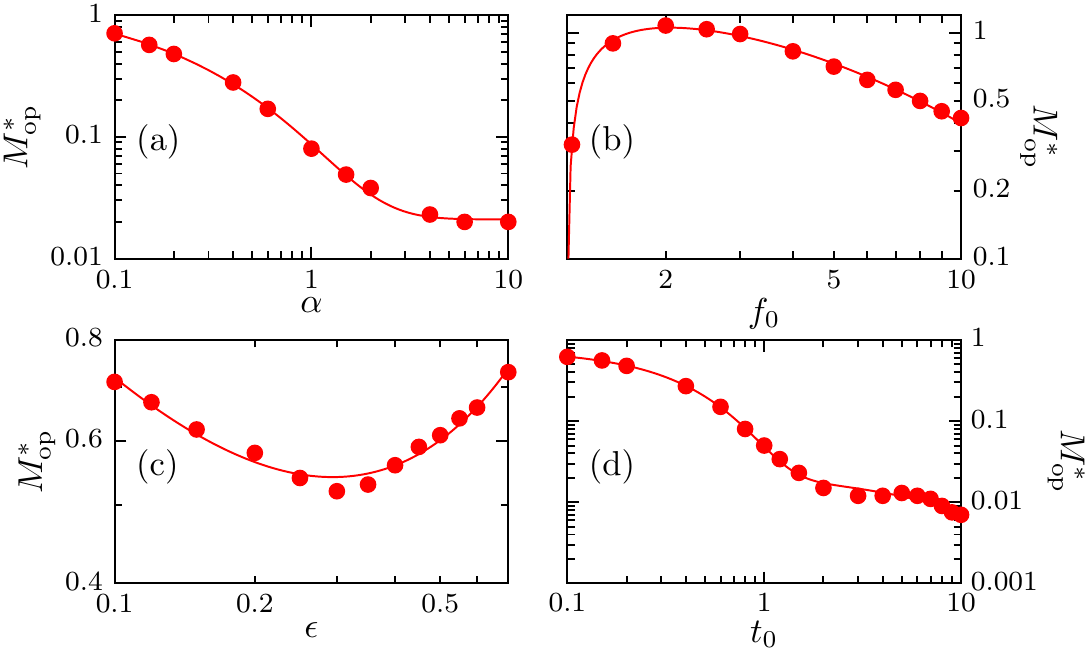}}
\caption{Dependence of $M^{*}_{\rm op}$ on the rotational diffusion rate $\alpha$ (a), self-propulsion force $f_0$ (b), aspect ratio of the channel $\epsilon$ (c), and active torque $t_0$ (d).
The moment of inertia $I^*=0.001$. }
\label{fig:Optimal_mass}
\end{figure}

The dependence of $M^{*}_{\rm op}$ on the rotational diffusion rate $\alpha$, self-propulsion force $f_0$, aspect ratio of the channel $\epsilon$, and active torque $t_0$ is shown in Fig.~\ref{fig:Optimal_mass}. For fixed values of $f_0$, $\epsilon$, and $t_0$, $M^{*}_{\rm op}$ decreases monotonically with $\alpha$ (panel (a)). From panel (b), for fixed values of $\alpha$, $\epsilon$, and $t_0$, $M^{*}_{\rm op}$ varies nonmonotonically with $f_0$ with the appearance of a peak. We observe that $M^{*}_{\rm op}$ is weakly dependent on $\epsilon$ (panel (c), see change in the ordinate scale); for fixed values of $\alpha$, $f_0$, and $t_0$, $M^{*}_{\rm op}$ has a minimum. Lastly, from panel (d), $M^{*}_{\rm op}$ monotonically decreases with $t_0$ for fixed values of $\alpha$, $f_0$, and $\epsilon$.

\section{Remarks and Conclusion}\label{Sec: Conclusions}

This study of the rectification and diffusion of self-propelled particles in a two-dimensional asymmetric channel showed that the inclusion of inertia leads to several distinctive features. In particular, most of the particles accumulate at the channel walls with increasing particle mass, while the remaining particles are distributed in a funnel-shaped region about the principal axis of the channel. This effect leads to enhanced rectification of heavier particles. The presence of a maximum in the effective diffusion coefficient as a function of the mass is also a consequence of the accumulation of most of the particles at the channel walls.
Furthermore, for various parameter values, the average particle velocity has a maximum as a function of the mass, indicating that particles with an optimal mass $M^{*}_{\rm op}$ drift faster than other particles; hence, they can be sorted from a mixture with particles of different masses.

While the Langevin model~(\ref{Eq:Langevin1}) can describe a wide variety of physical systems whose active agents are powered by various mechanisms and are subject to either thermal or athermal noise~\cite{Lowen-per2020}, it is instructive to discuss possible experimental realizations of the rectification effects described above. The asymmetric channels we considered can be constructed by microprinting on a substrate, and the effects of inertia can be determined from measurements of the average velocity and effective diffusion coefficient~\cite{Mahmud_et_al@NP:2009, Holz_et_al@NL:2009, Matthias_Muller@Nature:2003}. From the results presented in the text, one can see that the effects of inertia described above will manifest themselves only for certain values of the system parameters, in particular, the particle mass, friction coefficients, diffusion constants, self-propulsion force, solvent viscosity, etc. The ability to control all of these parameters within desirable ranges places limits on the physical systems.

A class of systems that may be of interest in this context are aerosols~\cite{Rohde_et_al@OE:2022}, where diffusiophoresis has been used to separate micrometer-scale particles~\cite{Bakanov1987,Chernyak2001}. As an example, consider identically-sized active particles with radii $a \sim 200~ \mathrm{nm}$, mass $M \sim 10^{-15} ~\mathrm{kg}$, moment of inertia $I \sim  10^{-31} ~\mathrm{kg~m^2}$ in air at room temperature and pressure $p=10^4-10^5~\rm{Pa}$. The viscosity is given by $\eta \sim 10^{-5}~\mathrm{kg/(m~s)}$, independent of pressure, with translational and rotational friction coefficients, $\gamma_t \sim 10^{-11} ~\mathrm{kg/s}$ and $\gamma_r \sim 10^{-24} ~\mathrm{kg~m^2/s}$, respectively. The active force lies in the range $F_0 \sim 0.1- 1 ~\mathrm{pN}$, and the active torque is taken to be zero. The translational and rotational noise strengths are then determined by the values of $D_t$ and $D_r$, respectively. The ratchet channel parameters are $L = 10~\mu{\rm m}$, ${\rm w}_\mathrm{max} = 10~\mu{\rm m}$, and $\epsilon =0.1$.

For thermal noise, the Einstein relations hold, and $D_t=k_BT/\gamma_t \sim 10^{-9}~\mathrm{m^2/s}$ and $D_r=k_BT/\gamma_r \sim 10^{3}~\mathrm{s^{-1}}$. In the dimensionless units introduced earlier, we have $\tau \sim 0.1 ~{\rm s}$, $M^* \sim 10^{-3}$, $I^* \sim 10^{-6}$, $f_0 \sim 10^{2}-10^{3}$, and $\alpha \sim 10^2$. From these parameter values, we can see that the regime where inertial effects play a role cannot be accessed.

For athermal noise, take $D_t \sim 10^{-8}-10^{-6}~\mathrm{m^2/s}$ and $D_r\sim 10^{3}~\mathrm{s^{-1}}$. We then have $\tau \sim 10^{-2}-10^{-4} ~{\rm s}$, $M^* \sim 0.01-1$, $I^* \sim 10^{-5}-10^{-3}$, $f_0 \sim 100-0.1$, and $\alpha \sim 10-0.1$. The persistence length and P\'eclet number are given by $l_{p} = v_0/D_r \sim 10-100~\mu{\rm m}$ and $Pe = v_0/ \sqrt{D_t D_r} \sim 0.3-30$, respectively. Under these conditions, the interesting regime where inertial effects lead to an optimal mass can be reached, provided the system is driven by external noise sources.

These results could stimulate the development of strategies for controlling the diffusion of active particles in entropic ratchet systems. Moreover, since the rectification of particles strongly depends on their mass, the model could be used to design lab-on-a-chip devices and artificial channels for the mass-based separation of particles.

\section{Acknowledgment}

This work was supported in part by the Natural Sciences and Engineering Research Council (NSERC) of Canada and Compute Canada (\href{https://www.computecanada.ca/}{www.computecanada.ca}).

% This work was supported by the Indian Institute of Technology
% Kharagpur under the Grant No. IIT/SRIC/PHY/TAB/2015-16/114.

%\begin{thebibliography}{100}

%\section*{References}

\bibliography{APS} %You need to replace "rsc" on this line with the name of your .bib file

\begin{thebibliography}{66}
\bibitem[{not({\natexlab{b}})}]{note:largeM}
\bibinfo{note}{The strongly underdamped limit may not be accessed easily for
  many physical systems; however, to better illustrate the trends of the
  transport properties, we have included results for very large $M^{*}$ in the
  plots.}
\end{thebibliography}

%\bibliographystyle{APS} %the RSC's .bst file

%\end{thebibliography}

\end{document}